\documentclass[prd, preprint,nofootinbib,superscriptaddress,12pt]{revtex4-1}
\pdfoutput=1

\newcount\showkey \showkey=0  %%% (1 showkeys, 0 no show)
\newcount\draft \draft=1 %%% (1 show remarks, 0 no show)

\usepackage[dvipsnames]{xcolor}
\usepackage{tabularx}

\ifnum\showkey=1
  \usepackage{seqsplit}
  \usepackage[color]{showkeys}
    \definecolor{refkey}{rgb}{0.9, 0.43, 0.63}
    \definecolor{labelkey}{rgb}{0.59, 0.43, 0.63}
  \usepackage{xstring}
  
\fi

\usepackage[pdfusetitle,pdfborder={0 0 0}]{hyperref}
   \hypersetup{colorlinks,
               linkcolor={red!70!black},
               citecolor={green!40!black},
               urlcolor={blue!80!black}}
\usepackage{microtype}   % Finetuned typesetting control
\usepackage{amsmath}     % AMS Math package; equation environments, math addons
  \usepackage{amsthm}    % AMS theorem-like environments
  \usepackage{amssymb}   % AMS math symbols and styles like \mathbb/cal/frak/...
\usepackage{siunitx}     % Unified number/unit handling
\usepackage{booktabs}    % Better table control
\usepackage[capitalize]{cleveref}    % Smart referencing
  
  \crefname{section}{Sec.}{Secs.}
  \crefname{appendix}{App.}{Apps.}

\usepackage[shortlabels,inline]{enumitem}  % Customize list envronments
\usepackage{mathtools} % Fine-tune math typesetting
\usepackage{subdepth}    % Alignment for sub/superscripts

\usepackage{graphicx,multirow,subfigure}
\usepackage{float} 
\usepackage[T1]{fontenc}
\usepackage{physics}
\usepackage{bbold}
\usepackage{rotating} %landscape mode for pages
\setlist[enumerate,2]{leftmargin=0.45em}
\usepackage{comment}

\DeclareFontFamily{U}{mathx}{\hyphenchar\font45}
\DeclareFontShape{U}{mathx}{m}{n}{<-> mathx10}{}
\DeclareSymbolFont{mathx}{U}{mathx}{m}{n}
\DeclareMathAccent{\widebar}{0}{mathx}{"73}

\newcommand{\beq}{\begin{equation}}
\newcommand{\eeq}{\end{equation}}

\ifnum\draft=1
  \newcommand{\ygn}[1]{{\bf \color{orange} #1}}
 \else
  \newcommand{\ygn}[1]{}
  \newcommand{\JBn}[1]{}
  \newcommand{\MFn}[1]{}
  \newcommand{\sts}[1]{}
  \newcommand{\JZn}[1]{}
\fi

\definecolor{nicered}{rgb}{0.7,0.1,0.1}
\definecolor{nicegreen}{rgb}{0.1,0.5,0.1}
\hypersetup{colorlinks,citecolor= blue,linkcolor= nicered}

\newenvironment{Eqnarray}{\arraycolsep 0.14em\begin{eqnarray}}{\end{eqnarray}}
\def\beqa{\begin{Eqnarray}}
\def\eeqa{\end{Eqnarray}}
\newcommand{\no}{\nonumber}

\newcommand{\bea}{\begin{eqnarray}}
\newcommand{\eea}{\end{eqnarray}}

\def\lsim{\mathrel{\rlap{\lower4pt\hbox{\hskip1pt$\sim$}}
     \raise1pt\hbox{$<$}}}         %less than or approx. symbol
\def\gsim{\mathrel{\rlap{\lower4pt\hbox{\hskip1pt$\sim$}}
     \raise1pt\hbox{$>$}}}         %greater than or approx. symbol

\begin{document}

\title{Lessons from ATLAS and CMS measurements of Higgs boson decays to second generation fermions}

\author{Yosef Nir}
\email{yosef.nir@weizmann.ac.il}
\affiliation{Department of Particle Physics and Astrophysics, Weizmann Institute of Science, Rehovot 7610001, Israel}

\author{Purvaash Panduranghan Udhayashankar}
\email{ purvaash.panduranghan-udhayashankar@weizmann.ac.il}
\affiliation{Department of Particle Physics and Astrophysics, Weizmann Institute of Science, Rehovot 7610001, Israel}

\begin{abstract}
There is now experimental evidence for Higgs boson decay into a pair of muons, and significant constraints on the Higgs boson decay into a charm quark-antiquark pair. The data on Higgs boson decays into second generation fermions probes various extensions of the Standard Model. We analyze the implications for the Standard Model effective field theory (SMEFT), without and with minimal flavor violation (MFV), for two Higgs doublet models (2HDM) with natural flavor conservation (NFC), for models with vector-like fermions, and for specific models that predict significant modifications of the Yukawa couplings to the light generations. 
\end{abstract}

\maketitle
%%%%%%%%%%%%%%%%
%%%%%%%%%%%%%%%%

\section{Introduction}
The ATLAS and CMS collaborations measured the decays of the Higgs boson into a tau-lepton pair, $h\to\tau^+\tau^-$, and into bottom-quark pair, $h\to b\bar b$, and the Higgs boson production in association with top-quark pair, $pp\to t\bar th$ \cite{CMS:2022dwd,ATLAS:2022vkf}. These measurements constitute a discovery of the mechanism that gives masses to the third generation fermions, as well as the discovery of Yukawa interactions among elementary particles \cite{Nir:2020wxg}. More recently, the ATLAS and CMS experiments searched for the Higgs boson decays into second generation fermions. 

The conventional way of reporting these measurements is by normalizing the production cross-section $\sigma(i\to h)$ and the branching ratio of the decay mode ${\rm BR}(h\to f\bar f)$ to their Standard Model (SM) values:
\beq
\mu_{f\bar f}^i=\frac{\sigma(i\to h){\rm BR}(h\to f\bar f)}{\sigma(i\to h)^{\rm SM}{\rm BR}(h\to f\bar f)^{\rm SM}}.
\eeq
One can interpret such measurement in terms of the Yukawa coupling modifier,
\beq
\kappa_f=Y_f/Y_f^{\rm SM}.
\eeq
Assuming that the production cross-section is not modified by new physics, and that only a single Yukawa coupling $Y_f$ is modified, the relation is given by
\beq
\mu_{f\bar f}=\frac{\kappa_f^2}{1+{\rm BR}^{\rm SM}_{h\to f\bar f}(\kappa_f^2-1)}.
\eeq

For the $h\to\mu^+\mu^-$ decay, the experiments report \cite{CMS:2020xwi,CMS:2022dwd,ATLAS:2020fzp,ATLAS:2022vkf}
\beqa\label{eq:muonexp}
\mu_{\mu^+\mu^-}&=&1.21^{+0.45}_{-0.42}\ \ \  [{\rm CMS}],\\
\mu_{\mu^+\mu^-}&=&1.2\pm0.6\ \ [{\rm ATLAS}],\no
\eeqa
with the average result \cite{PDG:2022}
\beq\label{eq:mumumu}
\mu_{\mu^+\mu^-}=1.21\pm0.35.
\eeq
This can be translated into
\beq\label{eq:kappamuexp}
0.92\leq\kappa_\mu\leq1.25.
\eeq

For the $h\to c\bar c$ decay, the experiments report \cite{CMS:2022psv,ATLAS:2022ers}
\beqa
\mu_{c\bar c}&=&7.7^{+3.8}_{-3.5}\ \ \  [{\rm CMS}],\\
\mu_{c\bar c}^{Vh}&=&-9\pm15\ \ [{\rm ATLAS}],\no
\eeqa
which translate into the following 95\% C.L. bounds:
\beqa\label{eq:kappacexp}
&&1.1<|\kappa_c|<5.5\ \ [{\rm CMS}],\\
&&|\kappa_c/\kappa_b|<4.5\ \ \ [{\rm ATLAS}].\no
\eeqa
At $1\sigma$, the CMS measurement translates into
\beq\label{eq:kcexp}
\kappa_c=3.1\pm1.0.
\eeq

There are two important discoveries implicit in these measurements:
\begin{itemize}
\item The muon-related measurement by CMS in Eq. (\ref{eq:muonexp}) confirms that the dominant source of the muon mass is the Yukawa interaction with the field of the scalar $h$.
\item The charm-related measurement by ATLAS in Eq. (\ref{eq:kappacexp}) confirms that $Y_c/Y_b<1$, in accordance with the SM prediction that $Y_c/Y_b=m_c/m_b$.
\end{itemize}
The measurements probe, however, also models that go beyond the SM (BSM). In this work, we explore these implications.
 
The plan of this paper is as follows. Sections~\ref{sec:smeft}-\ref{sec:higdep} analyze the implications of the $\mu_{\mu^+\mu^-}$ measurements in various relevant BSM frameworks: The SMEFT in Section~\ref{sec:smeft}, vector-like leptons in Section~\ref{sec:vll}, two Higgs doublet models (2HDM) in Section~\ref{sec:2hdm}, the Yukawa-less first-two-generation model in Section~\ref{sec:yukles}, and the Higgs-dependent Yukawa couplings model in Section~\ref{sec:higdep}.   Section~\ref{sec:charm} studies the implications of $\mu_{c\bar c}$ for these models. Our conclusions are summarized in Section~\ref{sec:con}.

%%%%%%%%%%%%%
\section{The Standard Model effective field theory (SMEFT)}
\label{sec:smeft}
The SM effective field theory (SMEFT) is the most general presentation of all BSM frameworks where the new degrees of freedom are much heavier than the electroweak scale, $v\approx246$ GeV. The SMEFT Lagrangian has the following form:
\beq
{\cal L}_{\rm SMEFT}={\cal L}_{\rm SM}+\frac{1}{\Lambda}{\cal O}_{d=5}+\frac{1}{\Lambda^2}{\cal O}_{d=6}+\ldots.
\eeq
Here, ${\cal L}_{\rm SM}$ is the Lagrangian of the renormalizable SM. The operators ${\cal O}_{d=n}$ are made of products of SM fields, of overall mass dimension $n$, contracted to a singlet of the SM gauge group. The scale $\Lambda$ is the mass scale of the new degrees of freedom, and by assumption $\Lambda\gg v$.

\begin{figure}[ht]
 \begin{center}
\includegraphics[width=0.46\textwidth]{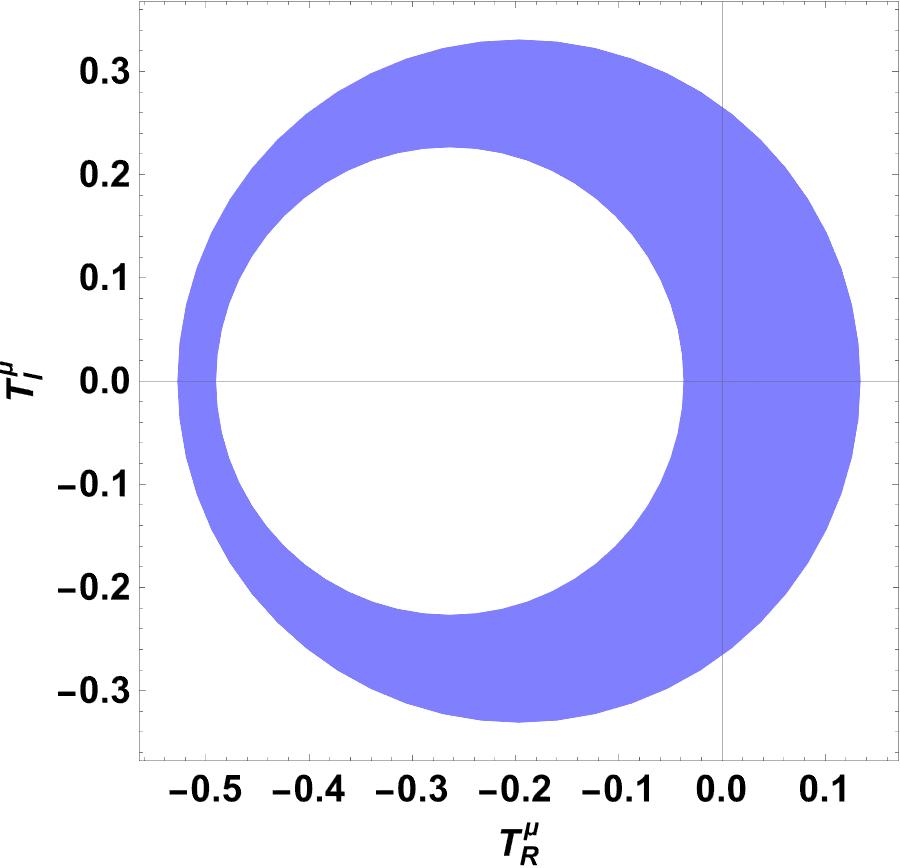}
  \caption{The allowed region in the $T_R^\mu-T_I^\mu$ plane.
  }
  \label{fig:trmu_timu}
 \end{center}
\end{figure}

Beyond the dimension-four Yukawa interaction of the SM, the muon mass and Yukawa coupling are modified by dimension-six terms:
\beq\label{eq:ldimsix}
{\cal L}_{\rm Yuk}^\mu=y_\mu\overline{L_{L\mu}}\Phi \mu_{R}+\frac{1}{\Lambda^2}(X_R^\mu+iX_I^\mu)|\Phi|^2\overline{L_{L\mu}}\Phi \mu_{R}+{\rm h.c.}.
\eeq
Here, $L_{L\mu}=(\nu_\mu\ \mu_L)^T$ is the left-handed $SU(2)_L$-doublet whose $T_3=-1/2$ member is the muon mass eigenstate, $\mu_R$ is the right-handed $SU(2)_L$-singlet muon mass eigenstate, and $\Phi=(0\ (v+h)/\sqrt{2})^T$ is the scalar $SU(2)_L$ doublet in the unitary gauge. Without loss of generality,we take the dimensionless coupling $y_\mu$ to be real.  It is convenient to define
\beq
T_{R,I}^\mu=\frac{v^2}{2\Lambda^2}\frac{X_{R,I}^\mu}{y_\mu}.
\eeq
Then, the mass and the effective muon Yukawa coupling are given by
\beqa
m_\mu&=&\frac{y_\mu v}{\sqrt{2}}\sqrt{(1+T_R^\mu)^2+(T_I^\mu)^2},\\
Y_\mu&=&\frac{y_\mu}{\sqrt2}\frac{1+4T_R^\mu+3(T_R^\mu)^2+3(T_I^\mu)^2+2iT_I^\mu}{\sqrt{(1+T_R^\mu)^2+(T_I^\mu)^2}}.
\eeqa
With our assumptions above that the production cross-section is not modified from the SM, that the muon Yukawa is the only one that is significnatly modified, neglecting the change in the total Higgs width and the effect of dimension-eight terms, we obtain \cite{Fuchs:2019ore}
\beq
\mu_{\mu^+\mu^-}=\frac{(1+3T_R^\mu)^2+9(T_I^\mu)^2}{(1+T_R^\mu)^2+(T_I^\mu)^2}.
\eeq
Thus, Eq. (\ref{eq:mumumu}) provides an allowed ring in the $T_R^\mu-T_I^\mu$ plane, as shown in Fig.~\ref{fig:trmu_timu}.

%%%%%%%%%%%%%%%%%%%%%%%%%%
%%%%%%%%%%%%%%%%%%%%%%%%%%%

Ref.~ \cite{Fuchs:2019ore} estimated the upper bounds on the CP-violating contributions of the dimension-six term to the baryon asymmetry of the universe,
\beq
\frac{|Y_B^{(\mu)}|}{2\times10^{-11}}\approx\frac{T_I^\mu}{(1+T_R^\mu)^2+(T_I^\mu)^2}\leq\frac{\sqrt{\mu^{\rm max}_{\mu^+\mu^-}}}{2},
\eeq
and to the electric dipole moment (EDM) of the electron:
\beq
\frac{|d_e^{(\mu)}|}{1\times10^{-30}\ e\ {\rm cm}}\approx\frac{T_I^\mu}{(1+T_R^\mu)^2+(T_I^\mu)^2}\leq\frac{\sqrt{\mu^{\rm max}_{\mu^+\mu^-}}}{2},
\eeq
where $\mu^{\rm max}_{\mu^+\mu^-}$ is the experimental upper bound on $\mu_{\mu^+\mu^-}$.
Given the observed value of the baryon asymmetry, $Y_B^{\rm obs}=(8.63\pm0.06)\times10^{-11}$ \cite{PDG:2022}, a complex Yukawa coupling of the muon can contribute at most (at the $1\sigma$ level)
\beq
|Y_B^{(\mu)}/Y_B^{\rm obs}|\lsim0.15.
\eeq
Given the experimental upper bound, $|d_e^{\rm max}|=4.1\times10^{-30}\ e\ {\rm cm}$ \cite{Roussy:2022cmp} (see also \cite{ACME:2018yjb}), a complex Yukawa of the muon is at most a factor of
\beq
|d_e^{(\mu)}/d_e^{\rm max}|\lsim0.15,
\eeq
below current experimental sensitivity.

The allowed range, which includes the SM value, $\mu_{\mu^+\mu^-}=1$, puts a lower bound on the scale of new physics that suppresses the dimension-six term. As long as  $\mu_{\mu^+\mu^-}=1$ is within the experimental range, the point $(T_R^\mu,T_I^\mu)=(-0.5,0)$, corresponding to $|T^\mu|=0.5$, is allowed, which implies 
\beq
\frac{\Lambda}{\sqrt{|X^\mu|}}\gsim\frac{v}{\sqrt{2 y_\mu^{\rm SM} |T^{\mu,{\rm max}}|}}\approx10\ {\rm TeV},
\eeq
where $y_\mu^{\rm SM}=\sqrt{2}m_\mu/v\approx6\times10^{-4}$. Since $|T|^2\propto1/\Lambda^4$, small deviations from $\mu_{\mu^+\mu^-}=1$ do not change this bound significantly.

In our analysis so far we assumed no special flavor structure for $X_{R,I}$. Imposing minimal flavor violation (MFV) implies that we expect $X_{R,I}^\mu={\cal O}(y_\mu)$ (and if we take MFV to imply that that Yukawa spurion is the only souce of CP violation, $X_I^\mu=0$). In this framework, $T_R^\mu={\cal O}(v^2/\Lambda^2)$, and consequently the bound on $\Lambda$ relaxes to $\Lambda\gsim v/\sqrt{(\mu^{\rm max}_{\mu^+\mu^-}-1)/2}\sim0.5\ {\rm TeV}$.

%%%%%%%%%%%
\section{Vector-like leptons}
\label{sec:vll}
Vector-like leptons affect the Higgs Yukawa couplings to pairs of the SM leptons if there are Yukawa terms involving a vector-like lepton and a SM lepton. There are five such vector-like representations, presented in Table \ref{tab:vllreps}. We follow here the notations and analysis of Ref.~\cite{Blum:2015rpa}.

%%%%%%%%%%%%%
 \begin{table}
\begin{center}
\begin{tabular}{|c|c|c|c| } \hline\hline
\rule{0pt}{1.2em}%
Names & $SU(2)_L\times U(1)_Y$ rep's &  $2M_X^2(c_{Y_e})_{\mu\mu}$ & $\kappa_\mu-1=$ \\[2pt]\hline\hline
\rule{0pt}{1.2em}%
$E+E^c$ & $(1)_{-1}+(1)_{+1}$ & $y_\mu|Y_{\mu_L E^c}|^2$ &  $-2\delta g_{A\mu}=-2\delta g_{V\mu}$ \\
$L+L^c$ & $(2)_{-1/2}+(2)_{+1/2}$ & $y_\mu|Y_{L \mu_R}|^2$ &  $-2\delta g_{A\mu}=+2\delta g_{V\mu}$ \\
$L^\prime+L^{\prime c}$ & $(2)_{-3/2}+(2)_{+3/2}$ & $y_\mu|Y_{L^\prime \mu_R}|^2$ &  $+2\delta g_{A\mu}=-2\delta g_{V\mu}$ \\
$T+T^c$ & $(3)_{0}+(3)_{0}$ & $2y_\mu|Y_{\mu_L T^c}|^2$ &  $+2\delta g_{A\mu}=+2\delta g_{V\mu}$ \\
$T^\prime+T^{\prime c}$ & $(3)_{-1}+(3)_{+1}$ & $y_\mu|Y_{\mu_L T^{\prime c}}|^2$ &  $-2\delta g_{A\mu}=-2\delta g_{V\mu}$ \\
\hline\hline
\end{tabular}
\caption{The leptonic vector-like representations. All fields are $SU(3)_C$ singlets.}
\label{tab:vllreps}
\end{center}
\end{table}
 %%%%%%%%%%%%%%

Taking the vector-like leptons to be much heavier than the electroweak scale, we integrate out these fields. This leads to, among others, the dimension-six term of Eq. (\ref{eq:ldimsix}), with a coefficient that we now denote by $(c_{Y_e})_{\mu\mu}$. In the presence of a single vector-like representation, the corresponding $(c_{Y_e})_{\mu\mu}$ is given in Table \ref{tab:vllreps}. The dimension-six terms contribute to the mass and the Yukawa coupling of the muon, and the relation between them is modified as follows:
\beq
Y_\mu=\frac{\sqrt{2}m_\mu}{v}+v^2(c_{Y_e})_{\mu\mu}.
\eeq

Each of the five vector-like representations modifies also the vector ($g_{V\mu}$) and axial-vector ($g_{A\mu}$)  $Z\mu^+\mu^-$ couplings. This modification is related to the modification of the Yukawa coupling as given in the fourth column ot Table~\ref{tab:vllreps} \cite{Blum:2015rpa}. We denote the modifications of the $Z$ couplings as follows:
\beq
\delta g_{V\mu}=g_{V\mu}-g_{V\mu}^{\rm SM},\ \ \ \delta g_{A\mu}=g_{A\mu}-g_{A\mu}^{\rm SM}.
\eeq
We use \cite{PDG:2022,ALEPH:2005ab}
\beqa
g_{V\mu}&=&-0.0367\pm0.0023,\ \ g_{V\mu}^{\rm SM}=-0.0371\pm0.0003\\
g_{A\mu}&=&-0.50120\pm0.00054,\ \ g_{A\mu}^{\rm SM}=-0.50127\pm0.00020,\no
\eeqa
and obtain
\beq\label{eq:delgav}
\delta g_{V\mu}=+0.0004\pm0.0023,\ \ \ \delta g_{A\mu}=+0.00007\pm0.00060.
\eeq
Combining  Table~\ref{tab:vllreps} and Eq.~(\ref{eq:delgav}), we learn that in these five models,
\beq
|\kappa_\mu-1|\lsim0.0013,
\eeq
two orders of magnitude below the experimental sensitivity, as reflected in Eq.~(\ref{eq:kappamuexp}).

A new ingredient is added in models that include two different vector-like representations in each, where there is a Yukawa coupling involving these two vector-like fields. These five models are presented in Table~\ref{tab:vllmodels}. 
%%%%%%%%%%%%%
 \begin{table}
\begin{center}
\begin{tabular}{|c|c|c| } \hline\hline
\rule{0pt}{1.2em}%
Model & Fields & $\kappa_\mu-1=$ \\[2pt]\hline\hline
\rule{0pt}{1.2em}%
LI & $L+L^c+E+E^c$ &  $-2\delta g_{A\mu}+X_\mu|Y_{L^c E}|$ \\
LII & $L^\prime+L^{\prime c}+E+E^c$ &  $-2\delta g_{V\mu}-X_\mu|Y_{L^{\prime c} E}|$  \\
LIII & $L+L^c+T^\prime+T^{\prime c}$ &  $-2\delta g_{V\mu}-X_\mu|Y_{L^c T^{\prime}}|$  \\
LIV & $L^\prime+L^{\prime c}+T^\prime+T^{\prime c}$ &  $-2\delta g_{A\mu}+X_\mu|Y_{L^{\prime c} T^\prime}|$  \\
LV & $L+L^c+T+T^c$ &  $+2\delta g_{V\mu}+X_\mu|Y_{L^c T}|$  \\
\hline\hline
\end{tabular}
\caption{Models with pairs of leptonic vector-like representations. $X_\mu$ is defined in Eq.~(\ref{eq:xmudef}).}
\label{tab:vllmodels}
\end{center}
\end{table}
 %%%%%%%%%%%%%%
In these models, the experimental ranges of $\delta g_{V\mu}$, $\delta g_{A\mu}$ and $(\kappa_\mu-1)$ constrain the combination $X_\mu Y_{F_1 F_2}$, where $Y_{F_1 F_2}$ is the Yukawa coupling between two vector-like leptons, and
\beq\label{eq:xmudef}
X_\mu=(v/m_\mu)\sqrt{2|\delta g_{V\mu}^2-\delta g_{A\mu}^2|}\exp(i\phi),
\eeq
and $\phi$ is a convention-independent phase.

Using the relations in Table \ref{tab:vllmodels}, and given that the allowed values of $\delta g_{V\mu}$ and $\delta g_{A_\mu}$ are about two orders of magnitude smaller than the bounds on $\kappa_\mu-1$, we obtain the following approximate bound:
\beq\label{eq:ydelg}
Y_{F_1 F_2}\sqrt{|\delta g_{V\mu}^2-\delta g_{A\mu}^2|}\lsim 3\times10^{-4}\times|\kappa_\mu-1|_{\rm max}\sim8\times10^{-5},
\eeq
where we use the upper bound on $\kappa_\mu-1$ from Eq.~(\ref{eq:kappamuexp}). Comparing Eq.~(\ref{eq:ydelg}) to Eq.~(\ref{eq:delgav}), we learn that, in the models defined in Table~\ref{tab:vllmodels}, for $Y_{F_1F_2}>{\cal O}(0.03)$, the $h\to \mu^+\mu^-$ measurement can be as constraining as measurements of  $Z\to\mu^+\mu^-$. 
 
%%%%%%%%%%%
\section{Two Higgs Doublet Models (2HDM)}
\label{sec:2hdm}
We consider two Higgs doublet models (2HDM) with natural flavor conservation (NFC). Common to all such models is the prediction that
\beq
\mu_{\mu^+\mu^-}=\mu_{\tau^+\tau^-}.
\eeq

For the $h\to\tau^+\tau^-$ decay, the experiments report \cite{CMS:2022dwd,ATLAS:2022yrq}
\beqa\label{eq:tauexp}
\mu_{\tau^+\tau^-}&=&0.85\pm0.10\ \ \  [{\rm CMS}],\\
\mu_{\tau^+\tau^-}&=&0.93^{+0.13}_{-0.12}\ \ [{\rm ATLAS}],\no
\eeqa
with the average result 
\beq\label{eq:mutautau}
\mu_{\tau^+\tau^-}=0.88\pm0.08,
\eeq
which translates into
\beq
\kappa_\tau=0.935\pm0.045.
\eeq
In this framework, we can combine Eq.~(\ref{eq:mumumu}) with Eq.~(\ref{eq:mutautau}) to obtain
\beq\label{eq:muellell}
\mu_{\ell^+\ell^-}=0.90\pm0.08.
\eeq
Comparing Eq.~(\ref{eq:mutautau}) to Eq.~(\ref{eq:muellell}), it is clear that the effect of the $\mu_{\mu^+\mu^-}$ measurement on the constraints on the model parameters is minor.

In 2HDM with NFC, there are three convenient bases. The interaction basis $(\Phi_1,\Phi_2)$, where each fermion sector ($u,d,e$) couples to one of the two doublets, the Higgs basis $(\Phi_M,\Phi_A)$, defined via $\langle\Phi_M\rangle=v$ and $\langle\Phi_A\rangle=0$, and the mass basis $(\Phi_h,\Phi_H)$, where the neutral CP-even members of the two doublets are the mass eigenstates $h$ and $H$. The angle $\beta$, defined via $\tan\beta=v_2/v_1$, rotates from the interaction basis to the Higgs basis, while the angle $\alpha$ rotates from the interaction basis to the mass basis.  

\begin{figure}[ht]
 \begin{center}
\includegraphics[width=0.46\textwidth]{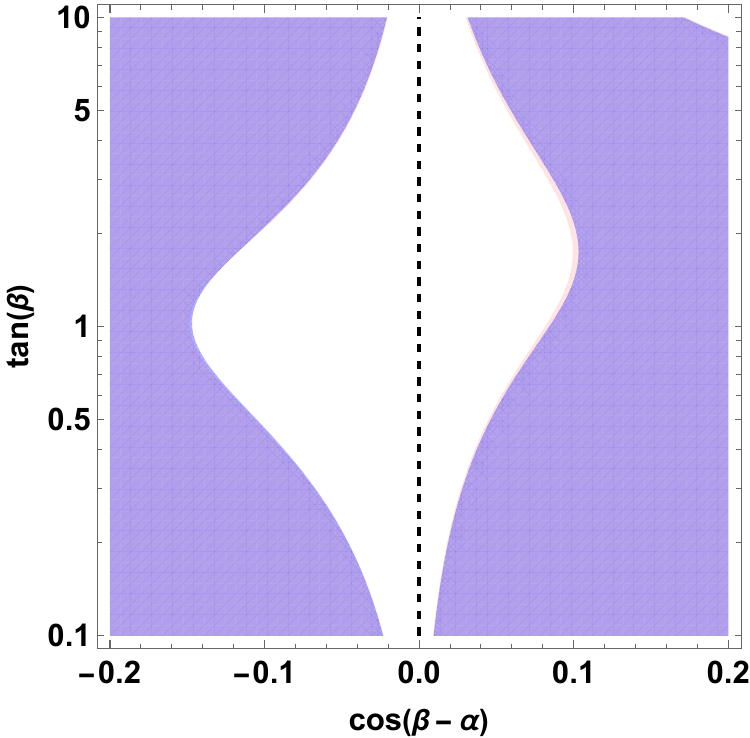}
\includegraphics[width=0.46\textwidth]{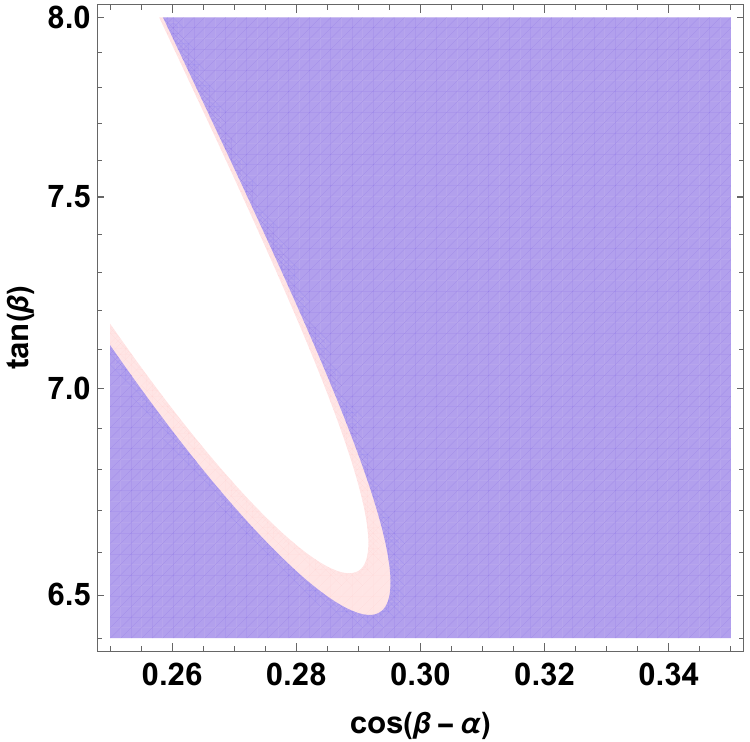}
  \caption{The excluded region in the $\cos(\beta-\alpha)-\tan\beta$ plane, in the lepton-specifc NFC model. The purple-shaded region does not include the $\mu_{\mu^+\mu^-}$ measurement, while the pink-shaded region is further excluded by this measurement. 
  }
  \label{fig:lepspe}
 \end{center}
\end{figure}

We demonstrate the smallness of the effect of the muon data in the lepton-specific type of NFC, where the quark ($u,d$) sectors couple to $\Phi_2$ and the lepton ($e$) sector couples to $\Phi_1$. Fig.~\ref{fig:lepspe} gives the excluded region in the $\cos(\beta-\alpha)-\tan\beta$ plane, without (purple) and with (purple plus pink) the muon constraint. We zoom in on the regions of $\cos(\beta-\alpha)\approx0$ (the decoupling limit) and $\cos(\beta+\alpha)\approx0$ (where the lepton Yukawa couplings are of opposite sign to the SM values).

%%%%%%%%%%%%%%%%%%%%%%%%%%
%%%%%%%%%%%%%%%%%%%%%%%%%%%

%%%%%%%%%%%%
\section{The Yukawa-less first-two-generation model}
\label{sec:yukles}
The Yukawa-less first-two-generation model of Ref.~\cite{Ghosh:2015gpa} was proposed, first, to demonstrate that it is possible that the mechanism that generates the masses of the first two generation fermions is different from the one that generates the masses of the third generation fermions, second, to provide an explanation for the lightness of the first two generations and, third, to demonstrate that it is possible that the two different mechanisms for fermion masses can be manifest in measurements of the Higgs boson decays to the light generations.

Consider a 2HDM. One of the two doublets, $\Phi_3$, couples dominantly to the third generation:
\beq
{\cal L}_3^Y=-\overline{L_{Li}}(Y_\tau\delta_{i3}\delta_{j3}+\epsilon^e_{ij})\Phi_3 E_{Rj}+{\rm h.c.},
\eeq
where $|\epsilon^e_{ij}|\ll1$ (and similarly in the quark sectors). The second doublet, $\Phi_{12}$, couples to only the first two generations:
\beq
{\cal L}_{12}^Y=-\sum_{i,j=1,2}\overline{L_{Li}}Y^{e12}_{ij}\Phi_{12} E_{Rj}+{\rm h.c.}.
\eeq
The scalar potential is given by
\beq
V=\mu_1^2|\Phi_{12}|^2+\mu_2^2|\Phi_3|^2+\mu^2(\Phi_{12}^\dagger\Phi_3+{\rm h.c.})+\lambda_1|\Phi_{12}|^4+\lambda_2|\Phi_3|^4.
\eeq
The VEVs, $\langle\Phi_3\rangle=v_3$ and $\langle\Phi_{12}\rangle=v_{12}$, are hierarchical:
\beq
\tan\beta=v_3/v_{12}\gg1,\ \ \ v_3^2+v_{12}^2=v^2.
\eeq

The scalar spectrum is given by
\beqa\label{eq:ylspectrum}
m_{H^\pm}^2&=&m_A^2=-\frac{\mu^2 v^2}{v_{12}v_3},\\
m^2_{h,H}&=&\lambda_1 v_{12}^2+\lambda_2 v_3^2-\frac{\mu^2 v^2}{2v_{12}v_3}\mp
\sqrt{\left(\lambda_1 v_{12}^2-\lambda_2 v_3^2+\frac{\mu^2(v_{12}^2-v_3^2)}{2v_{12}v_3}\right)^2+(\mu^2)^2},\no
\eeqa
and the rotation angle from the $(\Phi_{12},\Phi_3)$ basis to the $(\Phi_h,\Phi_H)$ basis, $\alpha$, is given by
\beq\label{eq:t2a}
\tan2\alpha=-\frac{2m_A^2 s_\beta c_\beta}{\sqrt{(m_H^2-m_h^2)^2-4m_A^4s_\beta^2c_\beta^2}}.
\eeq

Clearly, in this model, third generation fermions acquire their masses from interacting with the $\Phi_3$ field, while the first two generation fermions acquire their masses from interacting with $\Phi_{12}$. The question remains whether this situation will be manifest in the Yukawa couplings of the lighter, CP-even neutral scalar $h$. In the limit of very large $\mu$ and, in particular, $m_A^2\gg\lambda_2 v^2,\ \lambda_1 v_{12}^2$, we are in the decoupling limit, and $\kappa_f\simeq1$ for all fermions, including the light generations. Indeed, it is straightforward to show that, in this limit, Eq.~(\ref{eq:t2a}) implies that $s_\alpha\simeq -c_\beta$, and then $\kappa_\mu=-s_\alpha/c_\beta\simeq1$ and $\kappa_\tau=c_\alpha/s_\beta\simeq1$.

The situation is different in the case that $\lambda_1 v^2c_\beta^2\gg m_A^2,\ \lambda_2 v^2$ \cite{Ghosh:2015gpa}. Given the smallness of $c_\beta^2$, this hierarchy requires a very large $\lambda_1$. Ref.~\cite{Ghosh:2015gpa} argues that a natural embedding for this setup is a strongly coupled theory like technicolor. Eqs.~(\ref{eq:ylspectrum}) and (\ref{eq:t2a}) give
\beqa
m_h^2&=&2\lambda_2v^2,\\
m_H^2&=&2\lambda_1 v^2 c_\beta^2+m_A^2\gg m_A^2,\no\\
s_\alpha&=&-(m_A^2/m_H^2)c_\beta\ll c_\beta.\no
\eeqa
Consequently,
\beq\label{eq:kmuyule}
\kappa_\mu\approx m_A^2/m_H^2\ll1.
\eeq

The lower bound in Eq.~(\ref{eq:kappamuexp}) implies that, in contrast to Eq.~(\ref{eq:kmuyule}), $\kappa_\mu\not\ll1$ which, in the context of this model, implies that $\lambda_1 v^2c_\beta^2\not\gg m_A^2$, and the Yukawa-less first-two-generation model is excluded.

\begin{figure}[ht]
 \begin{center}
\includegraphics[width=0.46\textwidth]{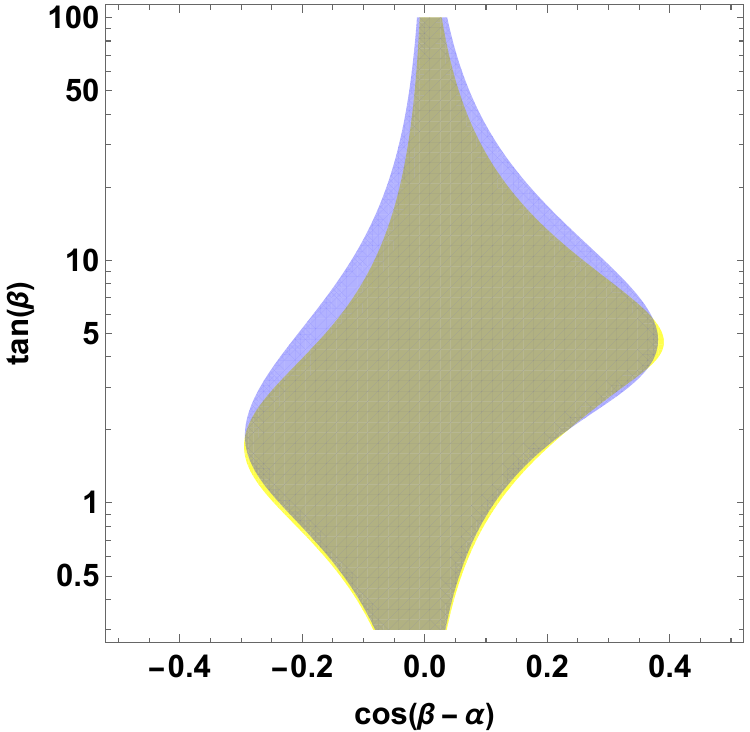}
  \caption{The allowed region in the $\cos(\beta-\alpha)-\tan\beta$ plane, in 1B 2HDM model of Ref.~\cite{Altmannshofer:2018bch}. When adding  the $\mu_{\mu^+\mu^-}$ measurement to all previous measurements of Higgs couplings, the previously allowed purple-shaded region is excluded, while the yellow-shaded region opens up. The grey region remains unchanged. 
  }
  \label{fig:twist1b}
 \end{center}
\end{figure}

Related models are the flavorful 2HDMs of Ref.~\cite{Altmannshofer:2018bch}. In their Type 1B 2HDM, the third generation fermions of all three sectors couple to $\Phi$, which has Yukawa couplings only in the $Y^f_{33}$ entries, while the first two generations couple to $\Phi^\prime$, which has small Yukawa couplings that break the $[U(2)]^5$ symmetry acting on the first two generations. Consequently, the measurement of $\mu_{\mu^+\mu^-}$ is the first to probe the couplings of $\Phi^\prime$. Yet, the effect of this measurement on the allowed region in the $\cos(\beta-\alpha)-\tan\beta$ plane is small, as can be seen in Fig.~\ref{fig:twist1b}.

%%%%%%%%%%%%%%%%%%%%%%%%%%
%%%%%%%%%%%%%%%%%%%%%%%%%%%

%%%%%%%%%%%%%%%%%%%%%
\section{Higgs dependent Yukawa couplings}
\label{sec:higdep}
The models of Refs.~\cite{Giudice:2008uua,Botella:2016krk} aim to explain the hierarchy in fermion masses by having the light fermion masses dominated by higher order temrs, {\it e.g.},
\beq
{\cal L}^\mu_{\rm Yuk}=c_{\mu\mu}\left(\frac{\Phi^\dagger\Phi}{\Lambda^2}\right)^{n_\mu}\overline{L_{L\mu}}\Phi\mu_R+{\rm h.c.},
\eeq
where $c_{\mu\mu}$ is a dimensionless coefficient of ${\cal O}(1)$, $\Lambda$ is a cut-off scale of ${\cal O}({\rm TeV})$, and $n_\mu$ is an integer $\geq1$. The muon mass and Yukawa coupling are given by
\beqa
m_\mu&=&c_{\mu\mu}\left(\frac{v}{\sqrt{2}\Lambda}\right)^{2n_\mu}\frac{v}{\sqrt{2}},\\
Y_\mu&=&c_{\mu\mu}(2n_\mu+1)\left(\frac{v}{\sqrt{2}\Lambda}\right)^{2n_\mu}.\no
\eeqa
Given that, experimentally, neither $\kappa_b\gg1$, nor $\kappa_\tau\gg1$, we assume that $n_b=n_\tau=0$. Then,
\beqa
\kappa_\mu&=&2n_\mu+1,\\
\mu_{\mu^+\mu^-}&=&\frac{(2n_\mu+1)^2}{1+{\rm BR}_{c,{\rm SM}}[(2n_c+1)^2-1]}.\no
\eeqa
Given that $m_c\gg m_\mu$, we assume that $n_c\leq n_\mu$. Thus, the minimal enhancement of $\mu_{\mu^+\mu^-}$ in this framework corresponds to $n_\mu=n_c=1$:
\beq
\mu_{\mu^+\mu^-}\geq\frac{9}{1+0.03\times8}\approx7.3.\no
\eeq
This value is clearly excluded by the experimental measurement of Eq.~(\ref{eq:mumumu}).

We conclude that the models of Higgs dependent Yukawa couplings of Refs.~\cite{Giudice:2008uua,Botella:2016krk} are excluded by the measurement of $\mu_{\mu^+\mu^-}$.

Another model where the Yukawa couplings are Higgs-dependent is the 2HDM model of Ref.~\cite{Bauer:2015fxa}. We postpone the discussion of this model to Section~\ref{sec:ccfirst2}, in order to include the top and charm data.

%%%%%%%%%%%%%%%%%
\section{Implications of $\mu_{c\bar c}$}
\label{sec:charm}
The measurement of $\mu_{c\bar c}$ suffers from low statistics. One should therefore be cautious about the SM value of $\kappa_c=1$ being outside the $95\%$ C.L. interval. Of course, if, with higher statistics, $\kappa_c=1$ will be exlcuded with high enough C.L., it will have striking implications. Here, however, we mention various implications for the BSM frameworks discussed in the previous sections, taking Eq.~(\ref{eq:kappacexp}) at face value.

%%%%%%%%
\subsection{The SMEFT}
We use notations similar to the discussion in section \ref{sec:smeft} (replacing $\mu$ with $c$).
There are several intriguing implications of the range of $\kappa_c$ in Eq.~(\ref{eq:kappacexp}):
\begin{itemize}
\item The range of $\kappa_c$ in Eq.~(\ref{eq:kappacexp}) corresponds to
\beq
1.3\ {\rm TeV}\lsim \frac{\Lambda}{\sqrt{|X^c|}}\lsim9.1\ {\rm TeV}.
\eeq
Thus, first, since $\kappa_c=1$ is excluded, we obtain also an upper bound on the scale $\Lambda$ and, second, the relevant new physics can be within the direct reach of the LHC.
\item The value of $\kappa_c=3$ is within the allowed range. This value corresponds to a situation where the contribution of the dimension-six term to $m_c$ dominates over the SM, dimension-four contribution.
\item Imposing MFV implies that we expect $X_{R,I}^c={\cal O}(y_c)$. In this case, the scale of new physics is required to be rather low, $\Lambda\lsim0.8\ {\rm TeV}$.
\end{itemize}

%%%%%%%%%%%
\subsection{Vector-like up-type quarks}
There are four representations of vector-like up-quarks that allow a Yukawa coupling to SM fields, presented in Table \ref{tab:vlureps}. We follow here the notations and analysis of Ref.~\cite{Blum:2015rpa}.

%%%%%%%%%%%%%
 \begin{table}
\begin{center}
\begin{tabular}{|c|c|c|c| } \hline\hline
\rule{0pt}{1.2em}%
Names & $SU(3)_C\times SU(2)_L\times U(1)_Y$ rep's &  $2M_X^2(c_{Y_u})_{cc}$ & $\kappa_c-1=$ \\[2pt]\hline\hline
\rule{0pt}{1.2em}%
$Q+Q^c$ & $(3,2)_{+1/6}+(\bar3,2)_{-1/6}$ & $y_c|Y_{Q c_R}|^2$ &  $+2\delta g_{Ac}=-2\delta g_{Vc}$ \\
$Q^{\prime\prime}+Q^{\prime\prime c}$ & $(3,2)_{+7/6}+(\bar3,2)_{-7/6}$ & $y_c|Y_{Q^{\prime\prime} c_R}|^2$ &  $+2\delta g_{Ac}=-2\delta g_{Vc}$ \\
$U+U^c$ & $(3,1)_{+2/3}+(\bar3,1)_{-2/3}$ & $y_c|Y_{c_L U^c}|^2$ &  $+2\delta g_{Ac}=+2\delta g_{Vc}$ \\
$U^\prime+U^{\prime c}$ & $(3,3)_{+2/3}+(\bar3,3)_{-2/3}$ & $y_c|Y_{c_L U^{\prime c}}|^2$ &  $-2\delta g_{Ac}=-2\delta g_{Vc}$ \\
\hline\hline
\end{tabular}
\caption{The up-quark vector-like representations.}
\label{tab:vlureps}
\end{center}
\end{table}
 %%%%%%%%%%%%%%

Each of the four vector-like representations modifies also the vector ($g_{Vc}$) and axial-vector ($g_{Ac}$)  $Zc\bar c$ couplings. This modification is related to the modification of the Yukawa coupling as given in the fourth column ot Table~\ref{tab:vlureps} \cite{Blum:2015rpa}. We denote the modifications of the $Z$ couplings as follows:
\beq
\delta g_{Vc}=g_{Vc}-g_{Vc}^{\rm SM},\ \ \ \delta g_{Ac}=g_{Ac}-g_{Ac}^{\rm SM}.
\eeq
We use \cite{ALEPH:2005ab}
\beqa
g_{Vc}&=&+0.1873\pm0.0070,\ \ g_{Vc}^{\rm SM}=+0.19204\pm0.00023\\
g_{Ac}&=&+0.5034\pm0.0053,\ \ g_{Ac}^{\rm SM}=+0.50144\pm0.00020,\no
\eeqa
and obtain
\beq\label{eq:delgavc}
\delta g_{Vc}=-0.0037\pm0.0070,\ \ \ \delta g_{Ac}=+0.0020\pm0.0053.
\eeq
Combining  Table~\ref{tab:vlureps} and Eq.~(\ref{eq:delgavc}), we learn that in these four models,
\beq
|\kappa_c-1|\lsim0.015,
\eeq
two orders of magnitude below the experimental sensitivity, as reflected in Eq.~(\ref{eq:kappacexp}).

A new ingredient is added in models that include two different vector-like representations in each, where there is a Yukawa coupling involving these two vector-like fields. These four models are presented in Table~\ref{tab:vlumodels}. 
%%%%%%%%%%%%%
 \begin{table}
\begin{center}
\begin{tabular}{|c|c|c| } \hline\hline
\rule{0pt}{1.2em}%
Model & Fields & $\kappa_c-1=$ \\[2pt]\hline\hline
\rule{0pt}{1.2em}%
UI & $Q+Q^c+U+Y^c$ &  $+2\delta g_{Ac}-X_c|Y_{Q^c U}|$ \\
UII & $Q^{\prime\prime}+Q^{\prime\prime c}+U+U^c$ &  $+2\delta g_{Vc}+X_c|Y_{Q^{\prime\prime U} E}|$  \\
UIII & $Q+Q^c+U^\prime+U^{\prime c}$ &  $-2\delta g_{Vc}+X_c|Y_{Q^c U^{\prime}}|$  \\
UIV & $Q^{\prime\prime}+Q^{\prime\prime c}+U^\prime+U^{\prime c}$ &  $-2\delta g_{Ac}-X_c|Y_{Q^{\prime\prime c} U^\prime}|$  \\
\hline\hline
\end{tabular}
\caption{Models with pairs of up-quark vector-like representations. $X_c$ is defined in Eq.~(\ref{eq:xcdef}).}
\label{tab:vlumodels}
\end{center}
\end{table}
 %%%%%%%%%%%%%%

In these models, the experimental ranges of $\delta g_{Vc}$, $\delta g_{Ac}$ and $(\kappa_c-1)$ constrain the combination $X_c Y_{F_1 F_2}$, where $Y_{F_1 F_2}$ is the Yukawa coupling of two vector-like up-quarks, and
\beq\label{eq:xcdef}
X_c=(v/m_c)\sqrt{2|\delta g_{Vc}^2-\delta g_{Ac}^2|}\exp(i\phi),
\eeq
and $\phi$ is a convention-independent phase.

Using the relations in Table \ref{tab:vlumodels}, and given that the allowed values of $\delta g_{Vc}$ and $\delta g_{A_c}$ are about two orders of magnitude smaller than the bounds on $\kappa_c-1$, we obtain the following approximate bound:
\beq\label{eq:ydelgc}
Y_{F_1 F_2}\sqrt{|\delta g_{Vc}^2-\delta g_{Ac}^2|}\lsim 4\times10^{-3}\times|\kappa_c-1|_{\rm max}\sim0.02,
\eeq
where we use the upper bound on $\kappa_c-1$ from Eq.~(\ref{eq:kappacexp}). Comparing Eq.~(\ref{eq:ydelgc}) to Eq.~(\ref{eq:delgavc}), we learn that, in the models defined in Table~\ref{tab:vlumodels}, for $Y_{F_1F_2}>{\cal O}(1)$, the $h\to c\bar c$ measurement can be as constraining as measurements of  $Z\to c\bar c$.

If $\kappa_c-1>0.1$ is established, then the models with vector-like up-quarks can explain the deviation from the SM with $Y_{F_1 F_2}\sqrt{|\delta g_{Vc}^2-\delta g_{Ac}^2|}\sim4\times10^{-4}$.
 
%%%%%%%%%%%%%%%%%%
\subsection{2HDM with NFC}
Common to all 2HDMs with NFC is the prediction that
\beq
\kappa_c=\kappa_t.
\eeq

For the $t\bar th$ production, the experiments report \cite{CMS:2022dwd,CMS:2020mpn,ATLAS:2017ztq}
\beqa\label{eq:texp}
\mu_{t\bar th}&=&0.94^{+0.20}_{-0.19}\ \ \  [{\rm CMS}],\\
\mu_{t\bar th}&=&1.2\pm0.3\ \ [{\rm ATLAS}],\no
\eeqa
with the average result \cite{PDG:2022}
\beq\label{eq:mutth}
\mu_{t\bar th}=1.10\pm0.18,
\eeq
which translates into
\beq\label{eq:kt}
\kappa_t=1.05\pm0.09.
\eeq
Taking the naive average of Eq.~(\ref{eq:kcexp}) and Eq.~(\ref{eq:kt}), we obtain
\beq\label{eq:ku}
\kappa_u=1.07\pm0.09.
\eeq
Comparing Eq.~(\ref{eq:kt}) to Eq.~(\ref{eq:ku}), it is clear that the effect of the $\mu_{c\bar c}$ measurement on the constraints on the model parameters is minor.

%%%%%%%%%%%%%%%%%
\subsection{Models with significantly modified second generation Yukawa couplings}
\label{sec:ccfirst2}
In the Yukawa-less first-two-generations model of Ref. \cite{Ghosh:2015gpa}, the Yukawa couplings of the Higgs boson to the first two generations are strongly suppressed compared to their SM values.
In the Higgs-dependent Yukawa couplings models of Refs. \cite{Giudice:2008uua,Botella:2016krk}, the Yukawa couplings of the Higgs boson to the first two generations are strongly enhanced compared to their SM values. As discussed in Sections \ref{sec:yukles} and \ref{sec:higdep}, these two models are excluded by the experimental value of $\mu_{\mu^+\mu^-}$ \cite{perez:2022}. Thus, we do not discuss them any further here.

Another model where the Yukawa couplings are Higgs-dependent is the 2HDM model of Ref.~\cite{Bauer:2015fxa}. In this model, we have
\beqa
\kappa_t&=&s_{\beta-\alpha}+c_{\beta-\alpha}/t_\beta,\\
\kappa_\tau&=&\kappa_b=3s_{\beta-\alpha}-2c_{\beta-\alpha}t_\beta+c_{\beta-\alpha}/t_\beta,\no\\
\kappa_c&=&3s_{\beta-\alpha}-c_{\beta-\alpha}t_\beta+2c_{\beta-\alpha}/t_\beta,\no\\
\kappa_\mu&=&5s_{\beta-\alpha}-3c_{\beta-\alpha}t_\beta+2c_{\beta-\alpha}/t_\beta.\no
\eeqa
We learn that $\kappa_c$ and $\kappa_\mu$ can be expressed in terms of $\kappa_t$ and $\kappa_\tau$,  and, consequently, the former are predicted in terms of the latter:
\beqa
\kappa_c&=&\frac12(\kappa_\tau+3\kappa_t)=2.0\pm0.2,\\
\kappa_\mu&=&\frac12(3\kappa_\tau+\kappa_t)=1.94\pm0.08.\no
\eeqa
While the value of $\kappa_c\sim2$ is within the currently allowed range, the value of $\kappa_\mu$ corresponds to $\mu_{\mu^+\mu^-}=3.5\pm0.3$, which is inconsistent with Eq.~(\ref{eq:mumumu}). We conclude that also this model is excluded.
 
%%%%%%%%%%%
\section{Conclusions}
\label{sec:con}
The ATLAS and CMS experiments at the LHC have been measuring branching ratios of the Higgs boson decays into second generation fermions, $h\to\mu^+\mu^-$ and $h\to c\bar c$. These measurements probe various BSM frameworks, where the SM relation between the Yukawa coupling and the mass, $y_f=\sqrt2 m_f/v$, is violated for second generation fermions. 

We obtain the following lessons from the $\mu_{\mu^+\mu^-}=1.21\pm0.35$ measurement:
\begin{itemize}
\item Models where there is a strong suppression of the Higgs boson Yukawa coupling to second generation fermions, such as the Yukawa-less first-two-generations model of Ref.~\cite{Ghosh:2015gpa}, are excluded.
\item  Models where there is a strong enhancement of the Higgs boson Yukawa coupling to second generation fermions, such as the Higgs-dependent Yukawa couplings models within either a single Higgs doublet \cite{Giudice:2008uua,Botella:2016krk} or two Higgs doublets \cite{Bauer:2015fxa}, are excluded.
\item In models where the SM fermion spectrum is extended by the addition of a single leptonic vector-like representation, the Higgs boson decay constraints are not competitive with the $Z$-boson decay constraints. In models with two vector-like representations, the two sets of constraints are complementary.
\item In 2HDMs with natural flavor conservation, constraints from $h\to\tau^+\tau^-$ are only slightly improved.
\item In the SMEFT, there is a lower bound on the scale that suppresses the relevant dimension-six terms of order 10 TeV. (The bound will not get stronger with better accuracy, as long as $\mu_{\mu^+\mu^-}=1$ remains within the allowed range.)
\end{itemize}

The CMS measurement of $\mu_{c\bar c}=7.7^{+3.8}_{-3.5}$ does not include the SM prediction, $\mu_{c\bar c}=1$, within the $95\%$ C.L. allowed range. Given the limited statistics, one should be cautious about its interpretation.  Taken at face value, we draw the following new lessons;
\begin{itemize}
\item Within the SMEFT, the scale of new physics should be within the ($1\sigma$) range of $1.3-9.1$ TeV.
\item In particular, models with two vector-like up-quark representations, can account from enhancement of the charm Yukawa coupling, if the Yukawa coupling involving the two vector-like representations is large enough, and with interesting implications for modifications of the $Zc\bar c$ coupling.
\item In 2HDM models with NFC, the constraints on the top Yukawa coupling from $\mu_{t\bar th}$ are in $\sim2\sigma$ tension with the $\mu_{c\bar c}$ constraint.
\end{itemize}

One of the most intriguing open questions related to the Higgs boson has been whether the second generation fermions acquire their masses through the Yukawa interaction with the field of the observed scalar $h$. For the muon, the $\mu_{\mu^+\mu^-}$ measurment now strongly supports that this is indeed the dominant mechanism. For the charm quark, the $\mu_{c\bar c}$ measurement still leaves open the possibility that this is not the case.

%%%%%%%%%%%%%%%%%
\section*{Acknowledgements}
YN is the Amos de-Shalit chair of theoretical physics, and is supported by grants from the Israel Science Foundation (grant number 1124/20), the United States-Israel Binational Science Foundation (BSF), Jerusalem, Israel (grant number 2018257), by the Minerva Foundation (with funding from the Federal Ministry for Education and Research), and by the Yeda-Sela (YeS) Center for Basic Research. 

%%%%%%%%%%%%%%%%%%%%%%%%

\end{document}